\documentclass{aa}  

\usepackage{graphicx}
\usepackage{txfonts}
\usepackage{lipsum}
\usepackage{subcaption}         
\usepackage{lscape}             
\usepackage{placeins}           

\usepackage{natbib}

\begin{document}

   \title{Type-III solar radio bursts with spike-like toppings}

   \subtitle{}

%
%
%

   \author{Shuwang Chang\inst{1,2}\email{shwchang@sdu.edu.cn}
        \and Chuanyang Li\inst{3}\corrauth{licy@sdu.edu.cn}
        \and Bing Wang\inst{2,4}\email{wbing@sdu.edu.cn}
        \and Guang Lu\inst{2,4}\email{guanglu@sdu.edu.cn}
        \and Zhao Wu\inst{2,4}\email{wuzhao@sdu.edu.cn}
        \and Fabao Yan\inst{1,2,5}\email{yanfabao2022@163.com}
        \and Hao Ning\inst{6}\email{haoning@sdu.edu.cn}
        \and Yao Chen\inst{2,4,6}\corrauth{yaochen@sdu.edu.cn}
        }

   \institute{School of Airspace Science and Engineering, Shandong University, Weihai 264209, China
   \and Laboratory for Electromagnetic Detection, Institute of Space Sciences, Shandong University, Weihai 264209, China
   \and College of Physics and Electronic Information, Dezhou University, Dezhou 253023, China
   \and School of Space Science and Technology, Shandong University, Weihai 264209, China
   \and  Shandong Key Laboratory of Intelligent Electronic Packaging Testing and Application, Shandong University, Weihai 264209, China
   \and Center for Integrated Research on Space Science, Astronomy, and Physics, Institute of Frontier and Interdisciplinary Science, Shandong University, Qingdao 266237, China}

   \date{Received April 15, 2026}

 
  \abstract
  {Spike-type~III burst pairs represent a distinct class of solar radio emissions in which clusters of spike-like bursts appear atop the high-frequency onset of type~III bursts. Using high time–frequency resolution data from the Chashan Broadband Solar radio spectrometer at meter wavelengths (CBSm), we present the largest statistical study to date of such events, comprising 502 spike–type~III pairs from 35 events recorded between November 2023 and October 2025. We find that spike-like clusters systematically precede their associated type~III bursts by 0.5--3~s in time ($\sim$87\% of pairs) and by 3--30~MHz in frequency ($\sim$80\%), a temporal and spectral offset that differs from earlier reports. The spike-like clusters exhibit diverse morphologies, including point-like, blob-like, drifting, and diffuse structures, with durations of $\sim$0.5--5~s and bandwidths of 15--150 MHz. Bi-directional drifting structures with rates of $\sim$20--100~MHz~s$^{-1}$ are observed, consistent with source motion both toward and away from the Sun. Furthermore, spike emission is predominantly strongly circularly polarized, with more than 64\% of clusters showing maximum polarization exceeding 0.6, in stark contrast to the generally weak polarization of type~III bursts. These findings point to an origin of the spike radiation in a multi-scale, inhomogeneous, and highly dynamic electron-acceleration region, providing novel observational constraints on the mechanisms underlying coherent solar radio bursts. }



   \maketitle

   \nolinenumbers

\section{Introduction}
Type III solar radio bursts are among the most common radio signatures of solar activity. They appear in the dynamic spectra as rapid drifting from high to low frequencies These bursts are widely interpreted as plasma emission exited by nonthermal electrons propagating along the open field lines in the flaring region, thereby exciting Langmuir waves. The observed radio emission is then produced via the wave-wave coupling of these Langmuir waves, generating radiation at the fundamental or harmonic frequency {\citep{wild1950, wild1963, melrose1980}}.

Near the type III burst onset, clusters of short-duration, narrowband spike-like structures occasionally appear (e.g., \citealt{benz1982, benz1996}, \citealt{benz2017}, \citealt{paesold2001}). According to earlier statistics {\citep{benz1982, benz1996}}, they are nearly synchronous in time and sense of polarization, yet distributed across different frequencies. Individual spikes are usually characterized by very short durations (typically tens to hundreds of milliseconds) and show little to no frequency drifting. Their full width at half maximum is about 4--6~MHz, corresponding to a relative bandwidth of $\sim$1\%--$2\%$ {\citep{benz1982, benz1996, paesold2001, benz2017, luo2021, musset2021}}.

Based on an analysis of 37 events (containing 60 spike clusters), \citet{benz1982} reported that: (1) spikes and type III bursts occur simultaneously, with spikes appearing above the type III burst showing only a small average frequency separation; (2) spikes usually appears in clusters with narrowband (bandwidth $\sim$4--$9$~MHz) and short durations ($\leq$0.05~s); (3) though not systematic analyzed, type III bursts associated with spikes tend to drift slowly; (4) while showing a wide range of polarization degrees, the average polarization of individual spikes is close to that of the associated type III burst, and spikes within a cluster usually share the same polarization sense.

\citet{benz1996} expanded the sample to 105 events, from which 13 were selected for a detailed analysis of the association between spikes and type III bursts, and they found: (1) the mean temporal offset between spikes and type III bursts at a common reference frequency is ⟨$\Delta t$⟩ = 0.03$\pm$ 0.04~s, i.e., almost no offset. Spikes tend to occur near or just above the type III starting frequency, with an average frequency separation $\bigtriangleup$$\nu$ (spike center frequency minus type III maximum frequency) of 23 $\pm$8~MHz. (2) Spikes typically last $\leq$0.1~s, have an average bandwidth of 9.9 $\pm$0.9~MHz, and usually form clusters. (3) Type III bursts accompanying these spikes drift relatively slowly, with a mean rate of about $-82 \pm 6$~MHz~s$^{-1}$. (4) Spikes are highly circularly polarized (mean 54 $\pm$ 11\%), while the associated type III bursts show weak polarization (mean 11 $\pm$ 2\%). About 60\% of the events share the same polarization sense between them.

\citet{paesold2001} combined spectral and imaging observations from the Nançay Radioheliograph (NRH). They found that: (1) the source position predicted by extrapolating the type III burst track to higher frequencies matches the observed spike location; (2) spikes are strongly polarized (up to 90--100\%), while the associated type III bursts show much weaker polarization ($\sim$10\%); (3) narrowband metric spikes are likely a direct radio signature of the particle acceleration region (or its immediate vicinity) in the solar corona.

Based on these statistical studies, \citet{benz1982, benz1996} proposed that spikes and type III bursts likely originate from the same plasma emission mechanism (both fundamental or both harmonic). The distinct properties of spikes—short duration, narrow bandwidth, and high polarization—point to a compact, strongly magnetized, or highly turbulent source region, possibly reflecting fragmented acceleration structures, current-sheet instabilities, or localized plasma turbulence (e.g., \citealt{karlicky2011}; \citealt{musset2021}). \citet{paesold2001} confirmed the close spatial association between spikes and type III bursts and thereby supported a common origin close to the electron acceleration region.

However, earlier instruments had limited time and frequency resolutions. For instance, the data used by \citet{benz1982, benz1996} typically had a time resolution of $\sim$0.1~s and a frequency resolution of $\sim$3~MHz, which prevented the detection of finer spectral structures and evolutionary details that has to wait solar spectrographs of higher resolution.

The Chashan Broadband Solar radio spectrometer at meter wavelengths (CBSm; \citealt{chang2024}) of the Chinese Meridian Project–Phase II (CMP II) operates in the 90--600~MHz band with high temporal resolution (0.84~ms), high frequency resolution (76.294~kHz), and high sensitivity ($\sim$1~SFU), making it well suited for observing fine spectral structures in meter‑wave solar radio bursts. Here we use the CBSm data to conduct a statistical study of spike–type~III burst pairs, focusing on their temporal, spectral, and polarization properties. Our goal is to obtain robust observational constraints and to discuss the physical origin of these events. The paper is organized as follows. Section~2 describes the instrument and the data. Section~3 presents an analysis of two representative events. Section~4 provides a statistical analysis of a sample of 35 events containing 502 spike–type~III pairs. Finally, Section~5 presents the conclusions and discussion.

\section{ Instrument and data}
\subsection{Instrument}
This study is primarily based on the data from CBSm. Its receiving system consists of a 12‑m parabolic dish, a dual‑linear‑polarized log‑periodic feed, and a high‑precision tracking platform. Polarization calibration of CBSm data has been available since 3 March 2025, with an uncertainty in polarization level of $\sim$<10\%. To verify the observation, we cross‑checked selected events with spectral data from the Daocheng Solar Radio Telescope (DART; \citealt{yan2023}), which is located $\sim$2100~km away from CBSm.

\subsection{Observations and data}
Since 10 November 2022, CBSm has accumulated thousands of solar metric radio bursts, including 65 type~II bursts and a substantial number of events exhibiting diverse fine spectral features. Data from CBSm have been used to investigate type~I, type~II, and type~IV radio bursts, as well as quasi periodic pulsations (e.g., \citealt{hou2023}; \citealt{yang2025}; \citealt{li2025}). The complete event catalogue is accessible online (\url{http://47.104.87.104/SRData/CBSm/}).

Till 6 February 2026, CBSm has recorded more than a hundred type~III burst groups accompanied by clusters of spike‑like topping (hereafter referred to as spike-like clusters). Each spike-like cluster and its accompanying type~III burst together constitute a spike–type~III pair. We selected 35 events with clearly defined morphology, yielding a total of 502 spike–type~III pairs for statistical analysis of their temporal and spectral properties. Of these, 309 pairs with available polarization data were examined to reveal their polarization behavior. Basic information on the selected events is given in Table~\ref{table:1}, including the date, observing interval, numbers of spike–type~III pairs per event, and the overall sense of polarization of the event.

\begin{table*}[ht!]
	\caption {List of events in the dataset and their key parameters} 
	\label{table:1} 
	\centering
	\begin{tabular}{c c c c c}
		\hline\hline             
		Number & Data & Time	& Count of spike-type & Polarization\\
		       &      &             & III bursts pairs    &             \\
		\hline
		1 & 2023-11-08      & 05:06:20--05:06:45 UT    & 8   & - \\
		2 & 2023-11-30      & 06:02:08--06:02:20 UT   & 9   & - \\
		3 & 2023-12-01      & 05:56:05--05:57:05 UT   & 12   & - \\
		4 & 2023-12-01      & 05:58:30--05:59:10 UT   & 18   & - \\
		5 & 2023-12-01      & 06:32:20--06:33:20 UT   & 9   & - \\
		6 & 2024-01-10      & 06:48:44--06:49:14 UT   & 10   & - \\
		7 & 2024-01-30      & 06:20:30--06:22:50 UT   & 41   & - \\
		8 & 2024-05-08      & 01:05:35--01:06:55 UT   & 21   & - \\
		9 & 2024-09-06      & 06:16:10--06:18:10 UT   & 15   & - \\
		10 & 2024-12-11      & 06:22:01--06:23:11 UT   & 19   & - \\
		11 & 2025-01-24      & 04:14:25--04:15:15 UT   & 8   & - \\
		12 & 2025-01-24      & 04:48:30--04:49:00 UT   & 12   & - \\
		13 & 2025-01-24      & 06:24:05--06:24:20 UT   & 4   & - \\
		14 & 2025-01-24      & 06:49:38--06:50:30 UT   & 7   & - \\
		15 & 2025-03-08      & 23:29:05--23:29:58 UT   & 20   & Right-handed \\
		16 & 2025-03-23      & 07:22:01--07:22:11 UT   & 11   & Right-handed \\
		17 & 2025-04-03      & 07:09:48--07:10:07 UT   & 8   & Right-handed \\
		18 & 2025-04-16      & 23:42:30--23:42:50 UT   & 8   & Left-handed \\
		19 & 2025-04-25      & 05:29:05--05:29:55 UT   & 20   & Right-handed \\
		20 & 2025-04-26      & 00:38:35--00:39:30 UT   & 11   & Negligible \\
		21 & 2025-04-27      & 07:59:05--07:59:40 UT   & 4   & Right-handed \\
		22 & 2025-04-27      & 08:34:00--08:34:50 UT   & 7   & Right-handed \\
		23 & 2025-04-28      & 01:36:30--01:38:00 UT   & 18   & Right-handed \\
		24 & 2025-04-28      & 01:38:20--01:39:00 UT   & 11   & Right-handed \\
		25 & 2025-04-28      & 02:54:30--02:55:10 UT   & 9   & Negligible \\
		26 & 2025-05-03      & 07:20:30--07:22:10 UT   & 40   & Left-handed \\
		27 & 2025-05-16      & 22:42:00--22:42:30 UT   & 11   & Left-handed \\
		28 & 2025-05-28      & 22:47:55--22:49:15 UT   & 30   & Left-handed \\
		29 & 2025-07-30      & 06:09:50--06:10:20 UT   & 15   & Right-handed \\
		30 & 2025-08-03      & 03:45:30--03:45:50 UT   & 11   & Right-handed \\
		31 & 2025-08-03      & 08:37:00--08:38:00 UT   & 14   & Right-handed \\
		32 & 2025-08-26      & 06:58:20--06:59:30 UT   & 16   & Left-handed \\
		33 & 2025-09-25      & 04:05:40--04:06:50 UT   & 24   & Right-handed \\
		34 & 2025-09-25      & 06:22:20--06:22:50 UT   & 10   & Right-handed \\
		35 & 2025-10-09      & 07:19:55--07:20:10 UT   & 11   & Right-handed \\
		\hline
	\end{tabular}
    \tablefoot{A dash (–) indicates that the measurement was not available for that event.}
\end{table*}

\begin{figure*}
	\centering
	\includegraphics[width=0.95\textwidth]{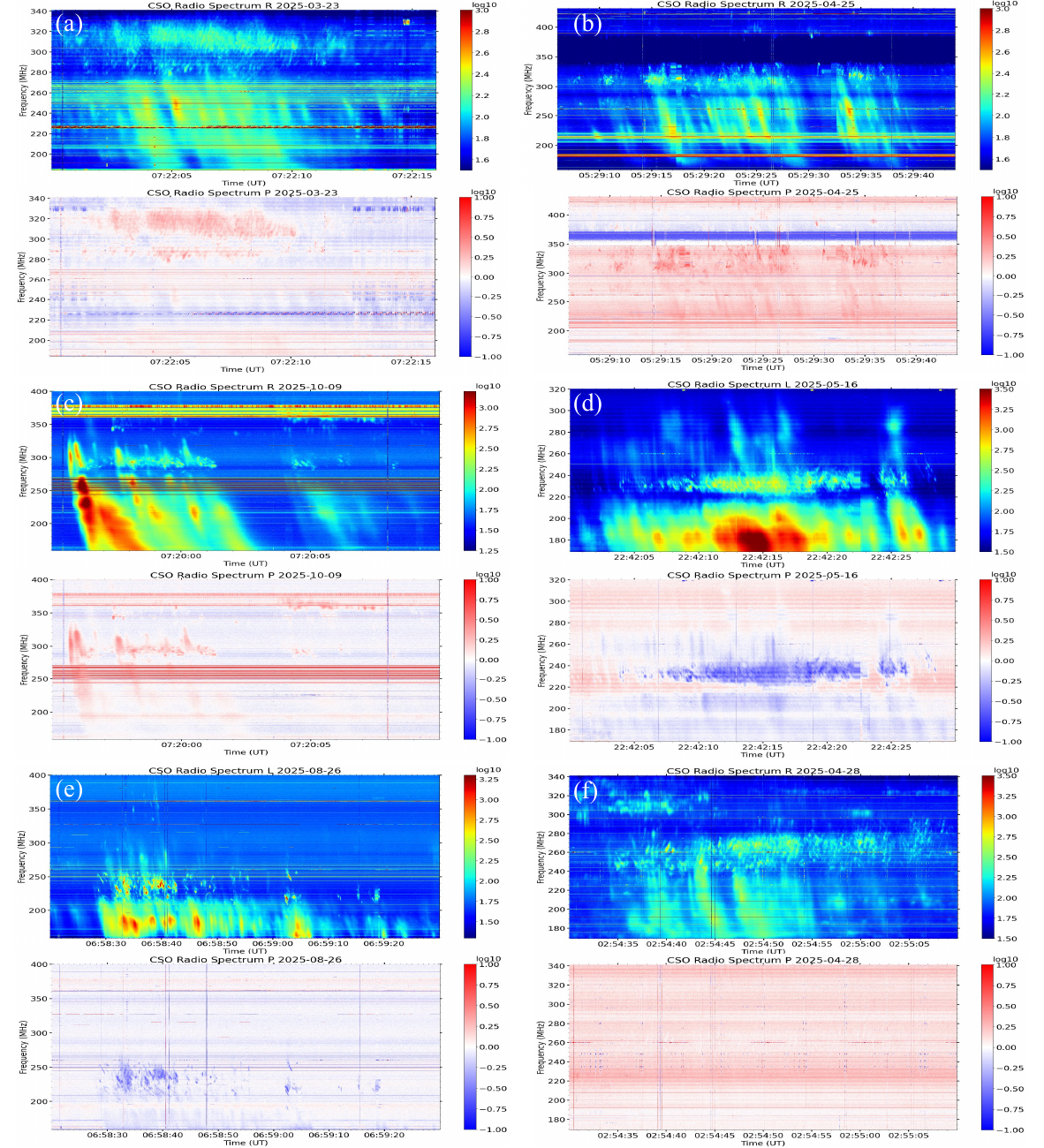}
	\caption{Examples of spike like clusters. Dynamic spectra and the corresponding polarization properties are shown for six representative events: (a) 23 March 2025 (07:22:00--07:22:16~UT), right-handed circular polarization; (b) 25 April 2025 (05:29:06--05:29:44~UT), right-handed; (c) 9 October 2025 (07:19:55--07:20:10~UT), right-handed; (d) 16 May 2025 (22:42:00--22:42:30~UT), left-handed; (e) 26 August 2025 (06:58:20--06:59:30~UT), left-handed; (f) 28 April 2025 (02:54:30--02:55:10~UT), no pronounced circular polarization.}
	\label{fig1}%
\end{figure*}

Figure~\ref{fig1} presents the dynamic spectra of six representative events. The frequency coverage, morphology, intensity, and degree of polarization of the type~III bursts and their spike-like clusters vary across the events. The count of spike–type~III pairs per event also differs; in each pair, the type~III burst and its associated spike-like cluster show a distinct offset in both time and frequency. The spikes in panels a--c display clear right‑handed circular polarization, those in panels~d and~e are left‑handed polarized, and the spikes in panel~f show no pronounced polarization.

\subsection{Identification of spike--type~III pairs}

We adopted the following multi-step procedure to separate spike-like clusters from type III bursts, particularly in complex cases.
\begin{enumerate}
	\item[(1)] We examined the polarization data. As noted in Section 4, in most events spike-like clusters exhibit high circular polarization (>0.6), while type~III bursts show low polarization (<0.4).
	\item[(2)] We examined the spectral morphology. Spike-like clusters appear as compact, narrowband features at higher frequencies, with little or no frequency drift. In contrast, type~III bursts show a characteristic fast, negative drift from high to low frequencies.
	\item[(3)] We checked temporal/spectral separation. In most spike—type~III pairs, the two components occupy distinct domains in time and/or frequency, providing a straightforward first-order identification.
	\item[(4)] For the few ambiguous cases where spike-like clusters and type~III bursts overlap, we separated them using an intensity threshold. If ambiguity persisted, we conducted a cross-correlation analysis (\citealt{benz1996}). We removed cases that remain ambiguous.
\end{enumerate}

\section{Analysis of two typical events}
This section presents a detailed analysis of two typical events observed on 18 March 2025 and 3 May 2025. A full statistical study of the entire sample will be presented in Section~4.

\subsection{Temporal, spectral, and polarization characteristics}
\begin{figure*}
	\centering
	\includegraphics[width=1.9\columnwidth]{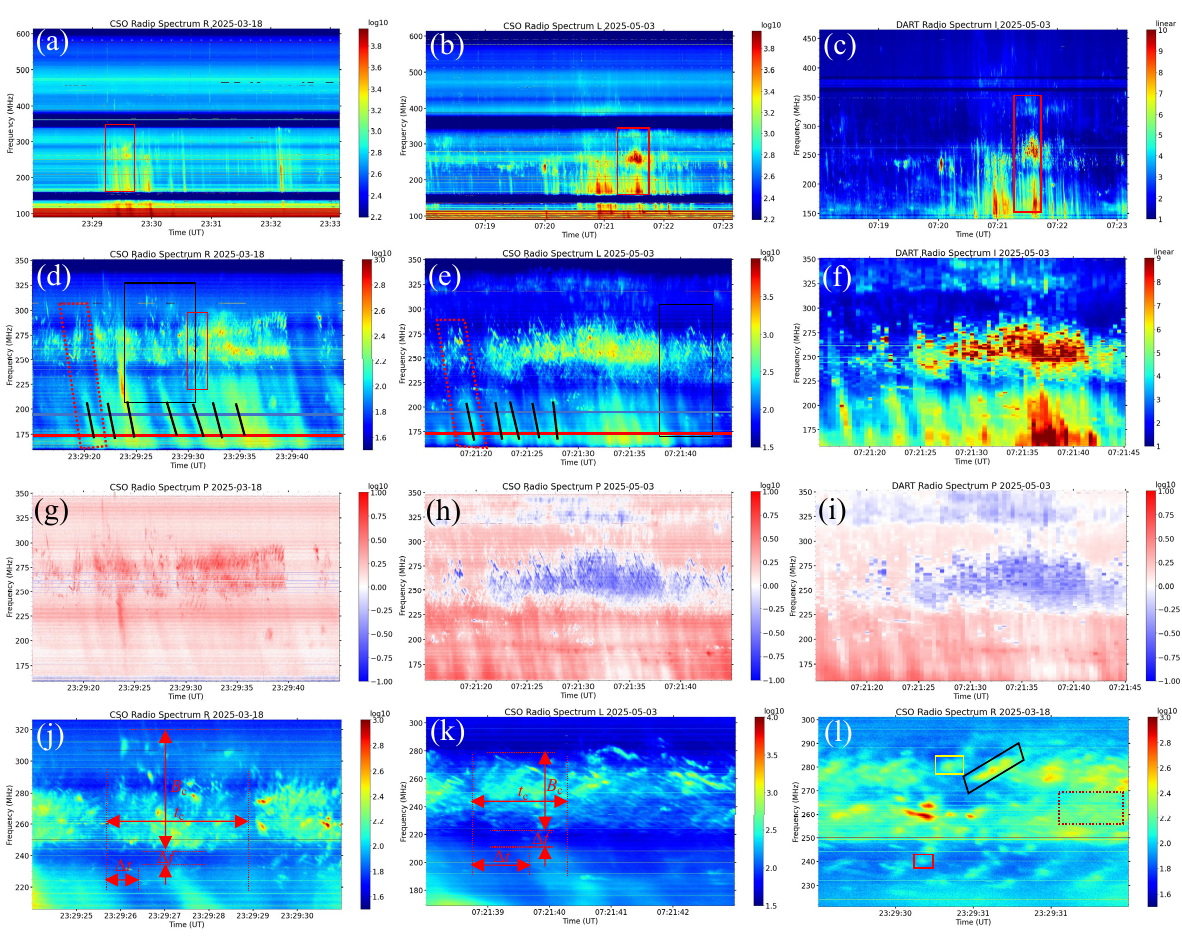}
	\caption{Dynamic spectra and circular polarization analysis for two representative events (18~March and 3~May 2025). (a,~b) Overview dynamic spectra of the 18~March (23:28:00--23:33:10~UT) and 3~May (07:18:00--07:23:10~UT) events. (c) Simultaneous overview from DART for the 3~May event. (d–f) Zoomed views of the red boxed regions in (a–c), respectively. (g–i) Degree of circular polarization for the regions shown in (d–f). (j,~k) Close ups of a typical spike–type~III pair (black boxes in~d and~e) annotated with the temporal lead $\Delta t$, frequency separation $\Delta f$, duration \textbf{$t_c$}, and bandwidth \textbf{$B_c$} of the spike like topping cluster. (l) Example of spike like toppings with different morphologies from the red boxed regions in (d): red boxes for point like; yellow boxes for blob-like; black boxes for drifting; purple dashed boxes for diffuse.}
	\label{fig2}%
\end{figure*}

Figure~\ref{fig2} shows the dynamic spectra of the two radio bursts; for comparison, the simultaneous DART spectrum of the 3~May event is also included. Both spectra clearly show a sequence of type~III bursts, each accompanied at its high‑frequency onset by spike‑like clusters. These structures appear in pairs or groups, forming spike–type~III pairs (Figs.~\ref{fig2}d–f).

In each spike–type~III pair, the spike-like cluster precedes the type~III burst. We define the leading time $\Delta t$ as the difference between their start times which are measured to be $\sim$1--2~s. The spike-like cluster is also located near the high‑frequency end of the type~III burst. The parameter $\Delta f$ represents the difference between the lowest frequency of the spike-like cluster and the highest frequency of the type~III burst. The measured $\Delta f$ varies in a range of 10--20~MHz (Figs.~\ref{fig2}j,~k).

\begin{figure*}
	\centering
	\includegraphics[width=0.95\hsize]{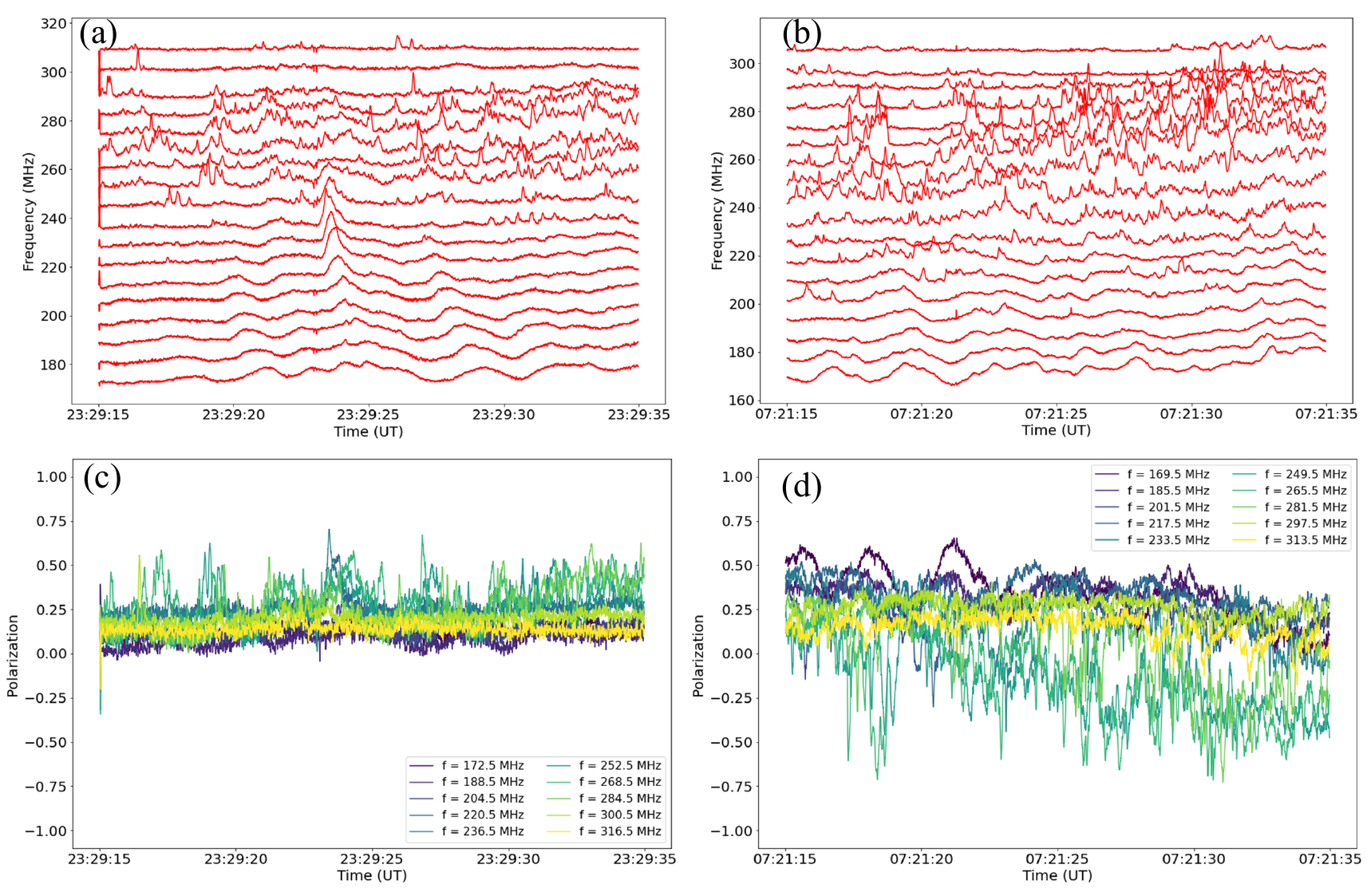}
	\caption{Time profiles of intensity and polarization at selected frequencies for the two events. (a) Normalized flux versus time for the 18~March 2025 event (23:29:15--23:29:35~UT) at 18 evenly spaced frequencies (172.5--316.5~MHz, step 8~MHz). (b) Same as (a) for the 3~May 2025 event (07:21:15--07:21:35~UT) with frequencies from 169.5 to 313.5~MHz. (c) Degree of circular polarization versus time for the same 18~March interval, shown at 10 evenly-spaced frequencies (172.5--316.5~MHz, step 16~MHz). (d) Same as (c) for the 3~May event, with frequencies from 169.5 to 313.5~MHz.}
	\label{fig3}%
\end{figure*}

Figures~\ref{fig3}a and~\ref{fig3}b present the normalized flux variations over 20~s at 18 selected frequencies for the two events, respectively. The plots clearly identify the spike–type~III pairs: spike-like clusters accompany the tops of the type~III bursts, which show negative frequency drifts of varying rates, and the spike-like clusters lead the type~III bursts in both time and frequency. Figures~3c and~3d show the corresponding temporal profiles of the degree of polarization at 10 frequencies. The spikes in the 18 March event display right‑handed circular polarization, whereas those on 3~May are left‑handed. In both events, the spike-like toppings and the associated type~III bursts share the same polarization sense, but the spike-like toppings are significantly more polarized. The simultaneous DART spectra in Figs.~\ref{fig2}c and~\ref{fig2}f are consistent with the CBSm results.

\subsection{Temporal and spectral characteristics of the spike-like clusters}
In both events, each type~III burst is accompanied by a spike-like cluster at its top (Figs.~\ref{fig2}d,~e,~j,~k). The spike-like clusters appear in succession. Some are separated by distinct time gaps, whereas others are closely spaced with less clear boundaries; however, each cluster retains a clear association with its corresponding type~III burst. For the 18~March event, the clusters have durations ($t_c$) of $\sim$1--4~s (mean $\approx$2.7~s), lie mostly in the 245--300~MHz band, exhibit bandwidths ($B_c$) of $\sim$45--65~MHz, and center frequencies that concentrate between 265 and~275~MHz. For the 3~May event, the clusters last $\sim$1.5--3.5~s (mean $\approx$2.4~s), occupy mainly the 230--300~MHz range, show bandwidths of $\sim$50--95~MHz, and have center frequencies that cluster between 235 and~255~MHz.

Individual spike-like structures exhibit diverse morphologies (see Fig.~\ref{fig2}(l)), such as point‑like (marked by red boxes), blob-like (yellow boxes), drifting (black parallelograms), and diffuse (purple dashed boxes) etc. The drifting structures move in opposite directions in both events. In Fig.~\ref{fig2}(l), a structure drifts from low to high frequency (positive drift) at $\sim$10--50~MHz~s$^{-1}$.

\subsection{Characteristics of the type~III bursts}
Figures~\ref{fig3}a and~\ref{fig3}b present the flux‑time profiles of the type~III bursts at selected frequencies. We determined the duration of each type~III burst from the full width of its intensity profile, and we extracted the time interval between successive bursts from the peak‑to‑peak separation. The burst intervals are $\sim$1--5~s, with a mean interval $\sim$2~s. A linear fit to the frequency–time data yields frequency drift rates of –75 to –25~MHz~s$^{-1}$ (mean $\approx$–41~MHz~s$^{-1}$) for the 18~March event and –100 to –40~MHz~s$^{-1}$ (mean $\approx$–67~MHz~s$^{-1}$) for the 3~May event (Figs.~\ref{fig2}d,~e).

The onset of the type~III burst often coincides with a short-lived intensification of the spike emission. In the representative cases shown in Fig.~\ref{fig3}, a sharp peak in the flux at higher frequency (dominated by spikes) is observed simultaneously with the start of the type~III burst. This is in line with the polarization data: the degree of circular polarization in the higher-frequency channels rapidly rises to values exceeding 0.6 at the same moment. This supports that this enhancement originates from spikes.

\section{Statistical analysis}
This section presents a statistical analysis of the 502 spike–type~III pairs identified from the 35 events.

\subsection{Statistics of the overall temporal and spectral characteristics}
\begin{figure}
	\centering
	\includegraphics[width=\hsize]{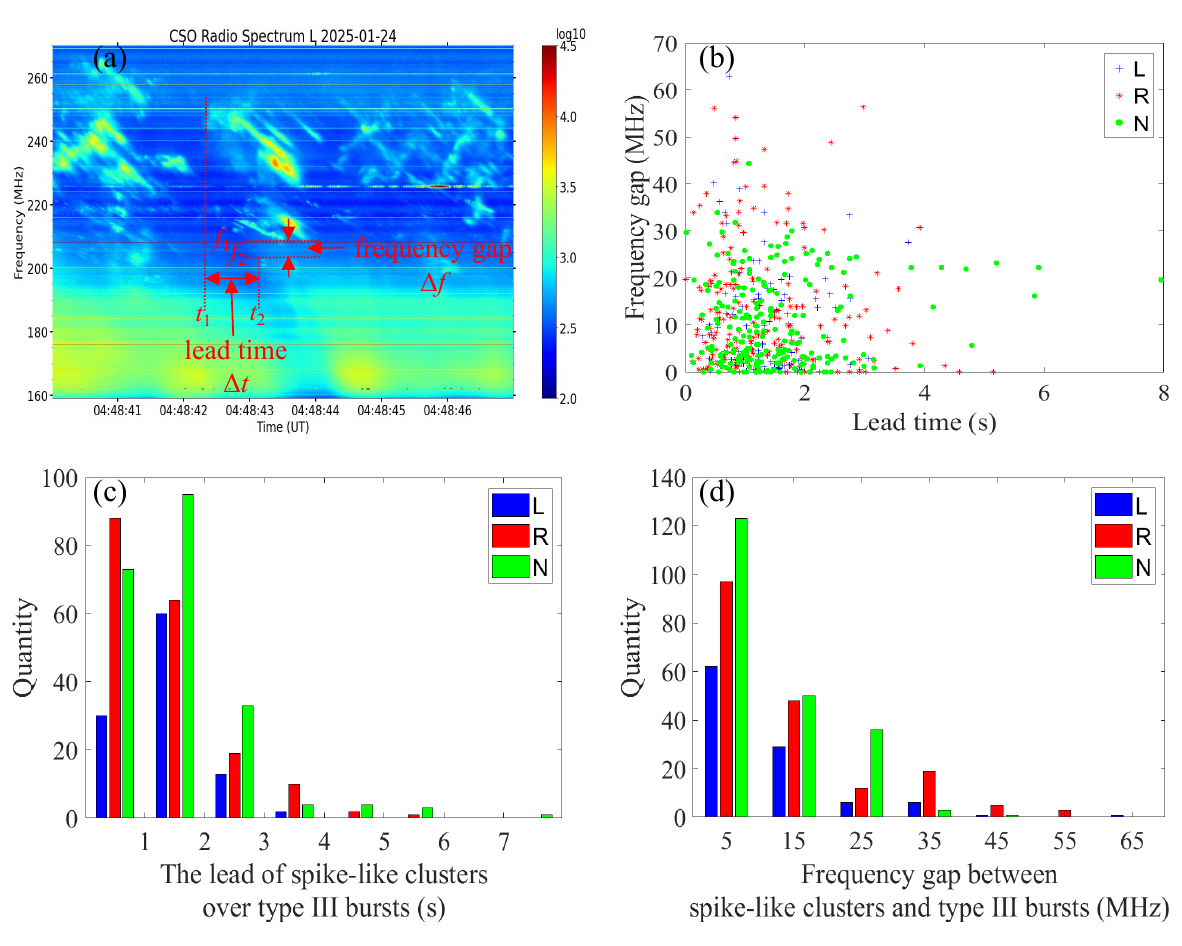}
	\caption{Temporal and spectral relationship between spike-like clusters and associated type~III bursts. (a) Schematic of a typical spike--type III pair, showing the lead time ($\Delta t$) and frequency separation ($\Delta f$). Here, $\Delta t = t_2 - t_1$, where $t_1$ is the start time of the spike-like cluster and $t_2$ is the start time of the type~III burst. $\Delta f = f_1 - f_2$, where $f_1$ is the lowest frequency of the spike-like cluster and $f_2$ is the highest frequency of the type~III burst. (b) Scatter plot of $\Delta t$ versus $\Delta f$ for all pairs. (c) Distribution of $\Delta t$. (d) Distribution of $\Delta f$. Data points in panels (b)--(d) are color-coded by the sense of polarization: red for right-handed (R), blue for left-handed (L), and green for unpolarized, nearly unpolarized, or without polarization data (N). This color coding scheme is also used in Fig.~\ref{fig5} and~\ref{fig7}.}
	\label{fig4}%
\end{figure}

A scatter plot in Fig.~\ref{fig4}b shows the lead time \textbf{($\Delta t$)} of the spike-like clusters relative to the type~III bursts versus the frequency separation \textbf{($\Delta f$)}. Here, $\Delta t$ is defined as the onset-time difference ($\Delta t = t_2 - t_1$), and $\Delta f$ as the difference between the lowest frequency of the spike-like cluster and the highest frequency of the type~III burst ( $\Delta f = f_1 - f_2$) (Fig.~\ref{fig4}a). The results show that $\sim$87\% of the spike-like clusters precede the associated type~III bursts by more than 0.5~s, and $\sim$83\% fall in the range of 0.5--3~s (Fig.~\ref{fig4}c). The frequency separation exceeds 3~MHz in $\sim$80\% of the pairs, with $\sim$63\% distributed between 3 and 30~MHz. Additionally, $\sim$59\% of the pairs have frequency separations greater than 5~MHz, and $\sim$52\% lie in the range of 5--30~MHz (Fig.~\ref{fig4}d).

\subsection{Temporal and spectral statistics of fine structures in spike-like toppings}
\begin{enumerate}
	\item[(1)] Duration of clusters
	
	Figure~\ref{fig5}a presents the duration distribution of 502 spike-like clusters, which range from 0.1 to 8~s, with $\sim$88.6\% between 0.5 and 5~s and $\sim$70.5\% between 1 and 4~s.
	
	\item[(2)] Frequency distribution 
	
    The spike-like clusters span the frequency range of 160–430~MHz, with  $\sim$84.5\% occurring between 180 and 370~MHz and  $\sim$66.1\% between 200 and 340~MHz. Bandwidths range from 10 to 220~MHz (Fig.~\ref{fig5}b), with  $\sim$89.2\% between 15 and 150~MHz and  $\sim$81.0\% between 20 and 130~MHz. The center frequency of each cluster, defined as the midpoint of its bandwidth, spans 170--400~MHz (Fig.~\ref{fig5}c), with  $\sim$91.4\% between 210 and 350~MHz and  $\sim$87.0\% between 220 and 330~MHz.

    \begin{figure}
	    \centering
    	\includegraphics[width=\hsize]{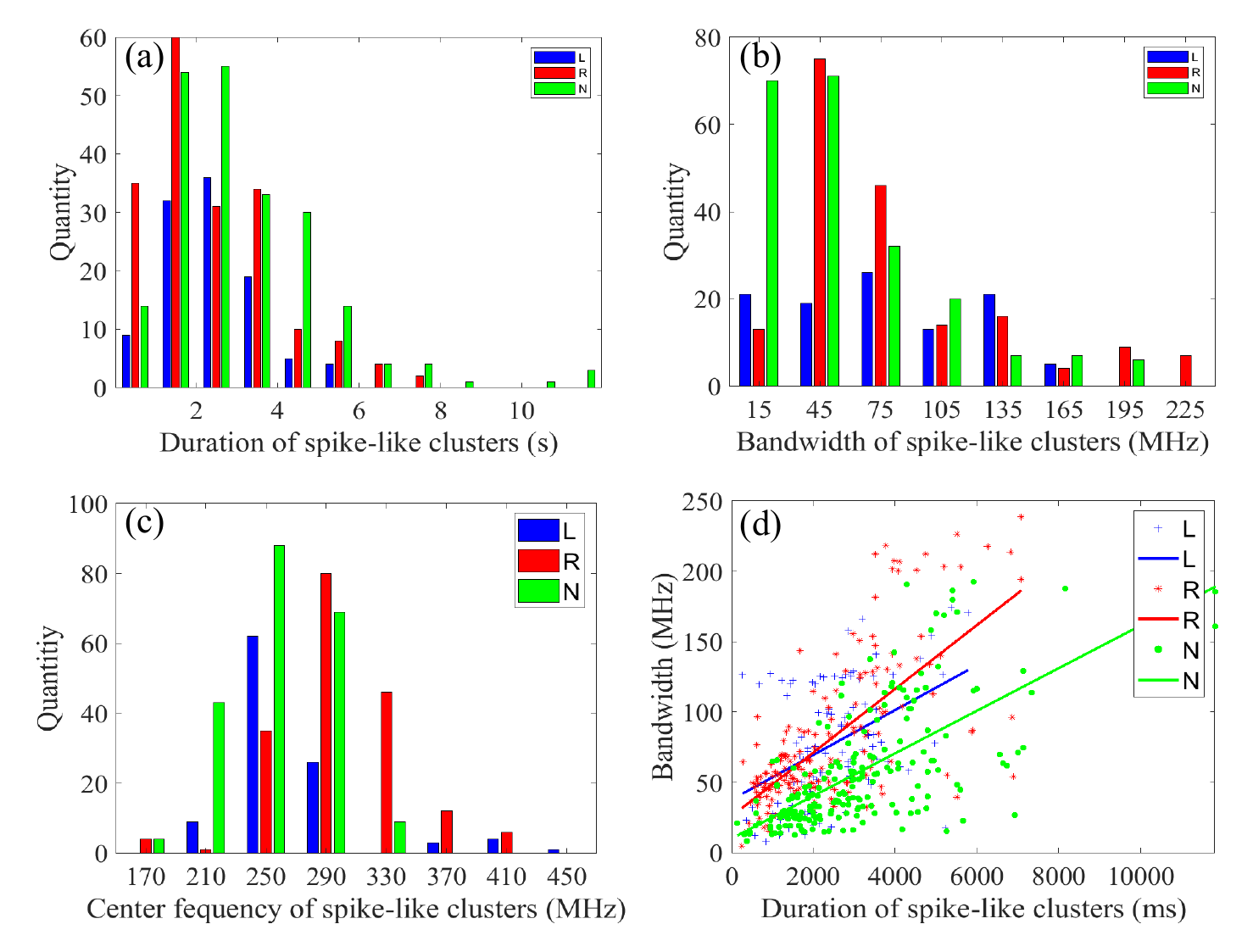}
	    \caption{Statistical distributions of spike-like clusters. (a) Duration. (b) Bandwidth. (c) Center frequency. (d) Duration vs. bandwidth.}
    	\label{fig5}%
    \end{figure}

    Figure~\ref{fig5}d shows a positive correlation between cluster duration and bandwidth. The correlation is observed for clusters with left handed, right handed, negligible, or unknown polarization, indicating that it is independent of polarization.
    		
	\item[(3)] Morphological characteristics 
	
	We measured the temporal and spectral parameters of the basic morphological types of the spike-like toppings: point‑like, blob-like, diffuse, and drifting structures. Overall, individual structures last $\sim$0.01--4~s, with most between 0.1 and 2~s, and have bandwidths of 1--70~MHz, predominantly 2--20~MHz. Specifically, point-like structures typically last 0.01--0.05~s with bandwidths of 1--5~MHz, and blob-like structures last 0.05--1.5~s and span $\sim$5–15~MHz.
	
	Among 35 events, 17 contain distinct drifting structures in their spike-like toppings, with 6 showing bidirectional drift and 11 showing unidirectional drift. Figure~\ref{fig6} presents examples from 9 of these events, covering both bidirectional (Figs.~\ref{fig6}a--l) and unidirectional (Figs.~\ref{fig6}m--r) drifts. Linear fits to these structures yield drift rates ranging from 20 to 130~MHz~s$^{-1}$ for positive drifts (frequency increasing) and –100 to –20~MHz~s$^{-1}$ for negative drifts (frequency decreasing) (see Table~\ref{table:2})
	
	\begin{figure*}
		\centering
		\includegraphics[width=0.9\textwidth]{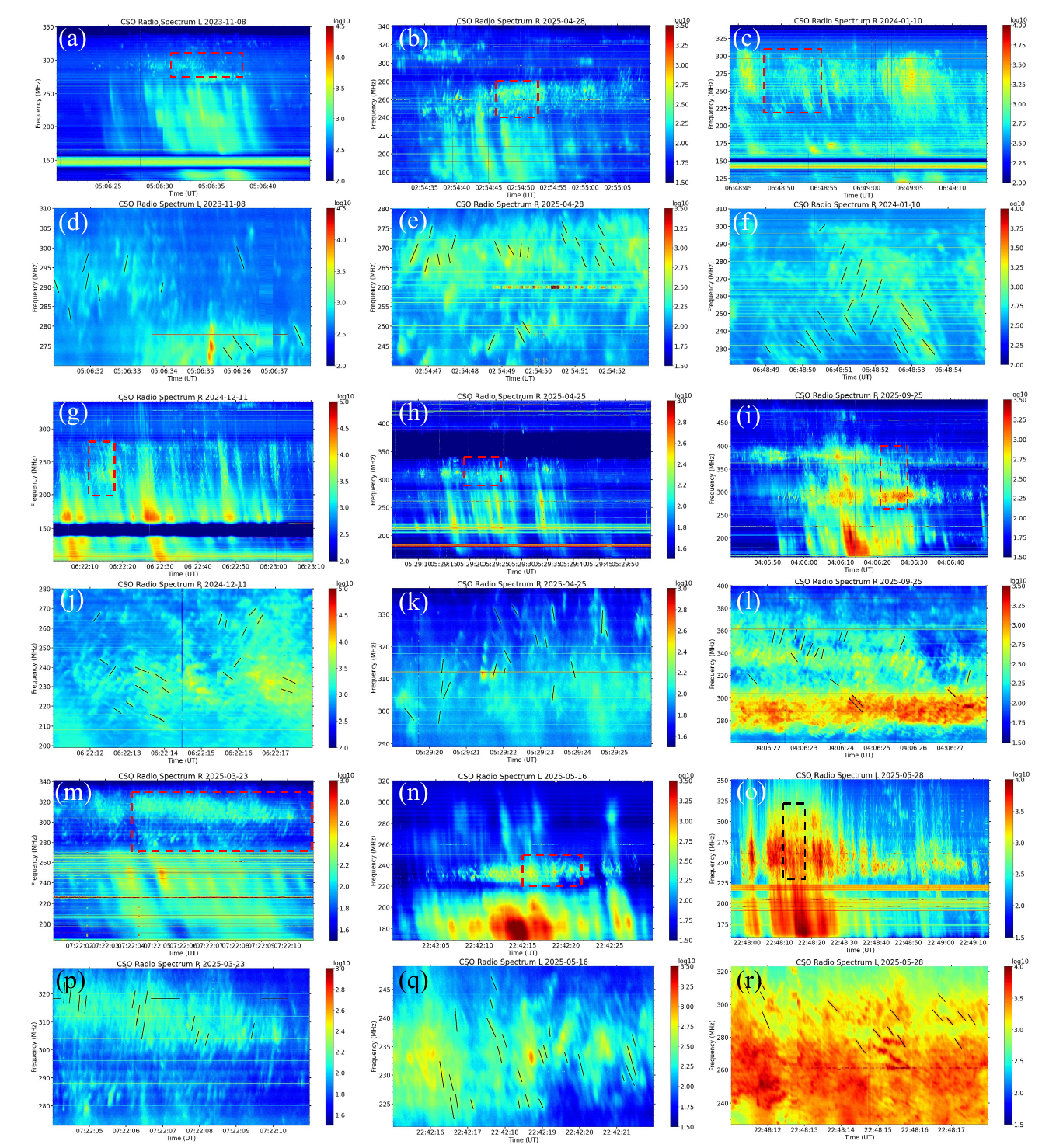}
		\caption{Examples of drifting structures in spike-like toppings. (a--c) Dynamic spectra observed on 8 November 2023 (05:06:20--05:06:45~UT), 28 April 2025 (02:54:30--02:55:10~UT), and 10 January 2024 (06:48:44--06:49:14~UT). (d--f) Expanded views of the red boxed regions in (a--c), respectively. (g--i) Spectra from 11 December 2024 (06:22:01--06:23:11~UT), 25 April 2025 (05:29:05--05:29:55~UT), and 25 September 2025 (04:05:40--04:06:50~UT). (j--l) Expanded views of the corresponding red boxed regions in (g--i). (m--o) Spectra from 23 March 2025 (07:22:01--07:22:11~UT), 16 May 2025 (22:42:00--22:42:30~UT), and 28 May 2025 (22:47:55--22:49:15~UT). (p--r) Expanded views of the corresponding red boxed regions in (m--o).}
		\label{fig6}%
	\end{figure*}
	
	\begin{table}[ht!]
		\caption{Drift rates of drifting structures of spike-like toppings} 
		\label{table:2}
		\centering
		\begin{tabular}{c c c c}
			\hline\hline
			Number & Event & \multicolumn{2}{c}{Drift rate (MHz~s$^{-1}$)} \\
			\cline{3-4}	
			&  & Positive & Negative \\
			\hline
			1 & Fig.~\ref{fig6}(a,~d) & 10--40 & $-40$--$-10$ \\ 
			2 & Fig.~\ref{fig6}(b,~e) & 20--110 & $-30$--$-10$ \\ 
			3 & Fig.~\ref{fig6}(c,~f) & 60--80 & $-40$--$-20$ \\ 
			4 & Fig.~\ref{fig6}(g,~j) & 2--12 & $-3$--$-1$ \\ 
			5 & Fig.~\ref{fig6}(h,~k) & 50--300 & $-130$--$-20$ \\ 
			6 & Fig.~\ref{fig6}(i,~l) & 30--130 & $-40$--$-20$ \\ 
			7 & Fig.~\ref{fig6}(m,~p) & 40--130 & -        \\ 
			8 & Fig.~\ref{fig6}(n,~q) & -      & $-65$--$-15$ \\ 
			9 & Fig.~\ref{fig6}(o,~r) & -      & $-100$--$-20$ \\ 
			\hline 
		\end{tabular}
		\tablefoot{A dash (–) indicates that the measurement was  not detected for that parameter.}
	\end{table}
		
\end{enumerate}

\subsection{Statistics of type~III burst characteristics}
\begin{enumerate}
	\item[(1)] Time intervals
	
	Time intervals $\Delta t$ between successive type\,III bursts are defined at a fixed frequency near each burst center as the peak-to-peak separation in intensity. The intervals range from $\sim$0.5 to $\sim$18~s, with $\sim$79\% between 1 and 5~s and a mean of $\sim$3~s.
	
	\item[(2)] Frequency Range 
	
	We measured the frequency ranges of the type~III bursts in the 502 spike–type~III pairs. The starting frequencies span $\sim$135--350~MHz, with a mean near 230~MHz; about 90\% exceed 190~MHz ($\sim$78\% are in 190--280~MHz) and roughly 80\% exceed 200~MHz ($\sim$68\% are in 200--280~MHz). In addition, the stopping frequencies of the type~III bursts exceed 90~MHz for about 67\% of the events.
	
	\item[(3)] Frequency drift rate 
	
	Frequency drift rates are measured for all 502 type~III bursts. Nearly all bursts show negative drift, ranging from –360 to –12~MHz~s$^{-1}$, with a mean of about –77~MHz~s$^{-1}$, and $\sim$85.6\% have drift rates between –100 and –20~MHz~s$^{-1}$ (Figs.~\ref{fig7}a,~b). There also exist N-type burst (\citealt{maxwell1958}; \citealt{aurass1997}; \citealt{kong2016}) with part of the burst drifting positively at $\sim$200~MHz~s$^{-1}$ for the event presented in Fig.~\ref{fig7}c.
	
\begin{figure}
	\centering
	\includegraphics[width=\hsize]{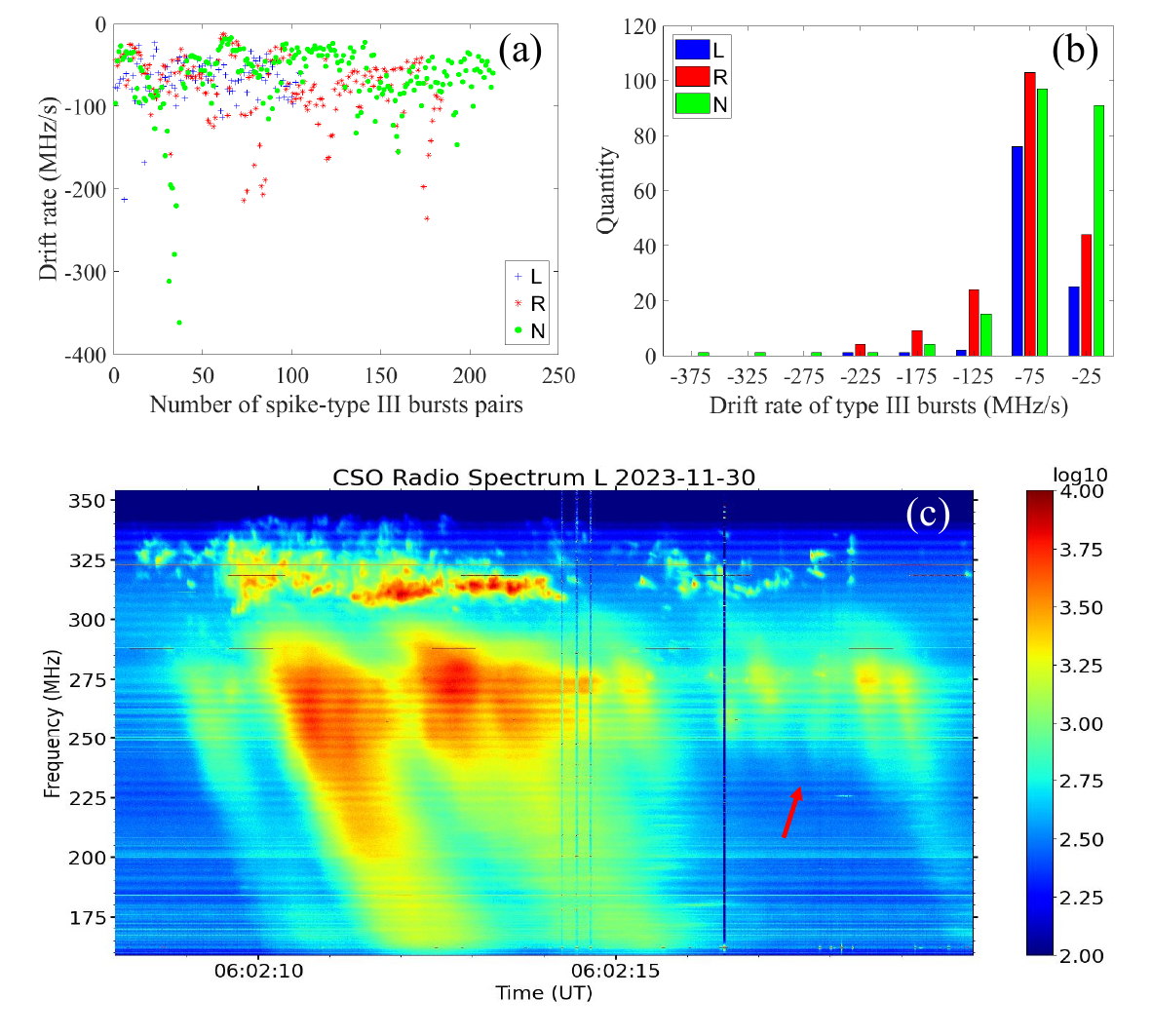}
	\caption{Frequency drift rates of type~III bursts, categorized by polarization state. (a) Scatter plot for 500 negatively drifting bursts, color coded as: red, right-handed polarization; blue, left-handed polarization; green, unpolarized, nearly unpolarized, or without polarization data. (b) Distribution of drift rates for the same sample. (c) Dynamic spectrum observed on 30 November 2023 (06:02:08--06:02:20~UT). The N‑shaped burst is marked by the red arrow.}
	\label{fig7}%
\end{figure}		
\end{enumerate}

\subsection{Statistics of polarization}
For each of the 309 pairs from 21 polarization-calibrated events, we measured the degree of circular polarization for both the spike-like cluster and its associated type~III burst, taking the maximum value in each component for statistical evaluation.

\begin{enumerate}
	\item[(1)] Polarization properties of the spike-like toppings
	
	Among 21 polarization-calibrated events, 14 of them show right-handed circular polarization, 5 left-handed, and 2 events show no significant polarization. Figure~\ref{fig8} and Table~\ref{table:3} present the distributions of maximum polarization for the spike-like clusters and the associated type~III bursts, including 184 right-handed and 105 left-handed pairs. In right-handed events, 54.9\% of clusters exhibit strong polarization (degree >0.6), compared to 80\% in left-handed events. Among the 289 left- and right-handed pairs, 287 clusters ($\sim$99.3\%) have a maximum polarization degree above 0.3, and 185 ($\sim$64\%) exceed 0.6.
	\begin{figure}
	\centering
	\includegraphics[width=\hsize]{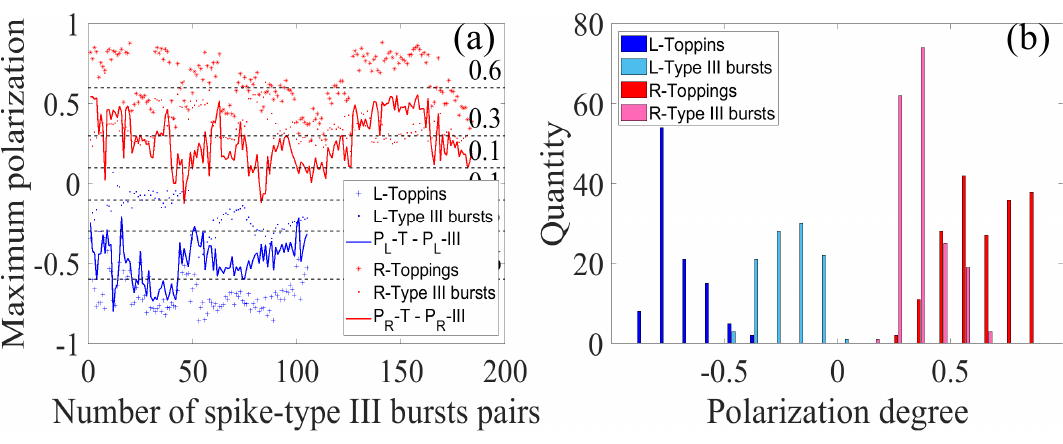}
	\caption{Maximum circular polarization: spike-like clusters versus associated type~III bursts (289 pairs). (a) Scatter plot of the maximum polarization of spike-like clusters against that of the bursts. Red/blue symbols indicate right-handed/left-handed polarization. The solid curve is a smoothed fit to the polarization difference of every pair. (b) Distribution of the polarization difference between spike-like clusters and type~III bursts.}
	\label{fig8}%
	\end{figure}

\begin{table*}[ht!]
	\caption{Polarization properties of spike–type~III pairs} 
	\label{table:3}
	\centering
	\begin{tabular}{c c c c c c}
		\hline\hline
		Polarization type & Object & \multicolumn{4}{c}{Intensity} \\
		\cline{3-6}	
		&                    & Negligible & Weak       & Moderate   & Strong  \\
		&                    & (<0.1)  & (0.1--0.3) & (0.3--0.6) & (>0.6) \\    
		\hline
		Right-handed & Spike-like toppings & 0       & 2         &  81         &  101  \\ 
		& Type III bursts    & 0       & 63        &  118        &  3    \\ 
		Left-handed  & Spike-like toppings & 0       & 0         &  21         &  84   \\ 
		& Type III bursts    & 22      & 58        &  24         &  1    \\ 
		\hline 
	\end{tabular}
\end{table*}
	
	\item[(2)] Polarization properties of type~III bursts 
	
	Type~III bursts show much weaker polarization than spike-like clusters, sometimes approaching the detection limit (Fig.~\ref{fig8}; Table~\ref{table:3}). In the 184 right-handed events, 63 bursts ($\sim$34.2\%) exhibit maximum polarization below 0.3, and 137 ($\sim$74.4\%) below 0.4. In the 105 left-handed events, 80 bursts ($\sim$76.2\%) remain below 0.3, and 100 ($\sim$95.2\%) below 0.4. In all 289 bursts, 49.5\% of them have maximum polarization below 0.3, and 82\% below 0.4, indicating that most type~III bursts are weakly or nearly unpolarized.
	
	We compared the maximum polarization of spike-like clusters with their associated type~III bursts. For the 184 right-handed pairs, $\sim$97.8\% of type~III bursts exhibit lower maximum polarization than the corresponding spike-like clusters, with the spike exceeding the burst by more than 0.2 in over 62\% of cases (red trend line in Fig.~\ref{fig8}a), while for the 105 left-handed pairs, type~III bursts also remain less polarized, with the difference exceeding 0.2 in every case (blue trend line in Fig.~\ref{fig8}a). Across all 289 clearly polarized pairs, spike-like clusters show maximum polarization at least 0.2 higher than the corresponding type~III burst in $\sim$75.8\% of pairs.
	
\end{enumerate}


\section{Conclusions and Discussion}
Using high time–frequency resolution CBSm data, we have conducted a statistical study of spike–type~III burst pairs using the largest-ever dataset of such bursts with 502 pairs of 35 events. Our findings are: (1) Spike-like clusters lead their associated type~III bursts in time by 0.5--3~s ($\sim$87\% of pairs) and in frequency by 3--30~MHz ($\sim$80\%). This temporal and spectral offset differs from the earlier analysis by \citet{benz1982, benz1996}; (2) Spike-like clusters exhibit diverse morphologies such as point-like, blob-like, uni-directional and bi-directional drifting, and diffuse forms; there exist bi-directional drifting structures with drifting rates of $\sim$20--100~MHz~s$^{-1}$; (3) many spike-like clusters are highly circularly polarized (maximum polarization >0.6 for $\sim$64\% of the sample), in stark contrast to the in-general weakly polarized type~III bursts (maximum polarization <0.4 for $\sim$82\%). In >75\% of the sample, the spike cluster’s polarization level exceeds that of the burst by >0.2. These results provide novel constraints on the origin of spike–type III~pairs.

Based on the high-resolution CBSm data, this work uncovers a diverse spectrum of morphologies within spike-like clusters, including point-like or blob-like structures lasting $\sim$0.01--1.5~s and with bandwidths of $\sim$1--15~MHz, implying a spatially compact source. Adopting energetic-electron speeds of 0.1--0.3c, the corresponding spatial scale can be as small as $\sim$10$^{3}$~km. A population of drifting structures is also detected, among which are newly identified bidirectional‑drifting features with drifting rates of $\sim$20--100~MHz~s$^{-1}$; such drifts are in line with source motion both toward and away from the Sun. Taken together, the variety of the spike emission, their systematic temporal lead ahead of the associated type~III bursts, and the well-defined frequency offset between the two—the observations point to an origin of the spike radiation in a multi-scale, inhomogeneous, and highly dynamic electron-acceleration region. The type~III bursts appear to be excited only after these energetic electrons escape the acceleration (spike-radiating) region and propagate outward for $\sim$0.5--3~s. In addition, we show that spike emission is predominantly strongly polarized, in contrast to the weak polarization that characterizes most type~III bursts. If both emissions arise via plasma emission, they must originate from distinctly different energetic-electron velocity distributions or coherent kinetic processes---a compelling possibility for follow-up study.

The study raises more questions than it addresses. For instance, why the two components have so different morphologies and polarization levels, and why they are temporally delayed by a certain time, and what these differences mean for the radiation mechanism? Are they both from the plasma emission mechanism, etc. Future work could combine hard X‑ray (HXR), soft X‑ray (SXR), and extreme‑ultraviolet (EUV) imaging to carry out multi‑wavelength analyses of spike–type~III pairs, including detailed case studies of events that exhibit clear morphology. In parallel, multi‑scale studies of electron acceleration and emission mechanisms are needed to better understand the observed differences in polarization and other properties between spikes and type~III bursts, and to clarify their origin.

As described in Section 3.3, spikes may get enhanced in intensity at the onset of type IIIs. This feature may agree with the near-zero mean delay reported by \citet{benz1996}.

\begin{acknowledgements}
      This work was supported by the National Natural Science Foundation of China (NNSFC) (Grant Nos. 12203031, 42127804 and 42374219), the National Key R\&D Program of China (Grant Nos. 2022YFF0503002, and 2022YFF0503000). We acknowledge the use of data from the Chinese Meridian Project. We thank the
      teams of CSO and DART for making their data available to us. The authors are grateful to the anonymous referee for the valuable comments.
\end{acknowledgements}

\end{document}